\documentclass[graybox]{svmult}

\usepackage{mathptmx}       
\usepackage{helvet}         
\usepackage{courier}        
\usepackage{type1cm}        
\usepackage{makeidx}         
\usepackage{graphicx}        
\usepackage{multicol}        
\usepackage[bottom]{footmisc}

\usepackage{subfig}

\makeindex             

\begin{document}

\title*{Empirical study of the influence of social groups in 
evacuation scenarios}
\author{Cornelia von Kr\"uchten, Frank M\"uller, Anton Svachiy, 
Oliver Wohak and Andreas Schadschneider}
\authorrunning{C. von Kr\"uchten, F. M\"uller, A. Svachiy, O. Wohak 
and A. Schadschneider}
\institute{Cornelia von Kr\"uchten \and Frank M\"uller \and Anton Svachiy 
\and Oliver Wohak \at Institut f\"ur Theoretische Physik, 
Universit\"at zu K\"oln, 50937 K\"oln, \email{cvk@thp.uni-koeln.de}
\and Andreas Schadschneider \at Institut f\"ur Physik und ihre Didaktik
and Institut f\"ur Theoretische Physik, Universit\"at zu K\"oln, 
50937 K\"oln, \email{as@thp.uni-koeln.de}}
%
%
\maketitle

\abstract*{The effects of social groups on pedestrian dynamics,
  especially in evacuation scenarios, have attracted some interest
  recently. However, due to the lack of reliable empirical data, most
  of the studies focussed on modelling aspects. It was shown that
  social groups can have a considerable effect, e.g. on evacuation
  times.  In order to test the model predictions we have performed
  laboratory experiments of evacuations with different types and sizes
  of the social groups. The experiments have been performed with
  pupils of different ages. Parameters that have been considered are
  (1) group size, (2) strength of intra-group interactions, and (3)
  composition of the groups (young adults, children, and mixtures).
  For all the experiments high-quality trajectories for all
  participants have been obtained using the PeTrack software.
  This allows for a detailed analysis of the group effects. One
  surprising observation is a decrease of the evacuation time with
  increasing group size.}

\abstract{The effects of social groups on pedestrian dynamics,
  especially in evacuation scenarios, have attracted some interest
  recently. However, due to the lack of reliable empirical data, most
  of the studies focussed on modelling aspects. It was shown that
  social groups can have a considerable effect, e.g. on evacuation
  times.  In order to test the model predictions we have performed
  laboratory experiments of evacuations with different types and sizes
  of the social groups. The experiments have been performed with
  pupils of different ages. Parameters that have been considered are
  (1) group size, (2) strength of intra-group interactions, and (3)
  composition of the groups (young adults, children, and mixtures).
  For all the experiments high-quality trajectories for all
  participants have been obtained using the PeTrack software.
  This allows for a detailed analysis of the group effects. One
  surprising observation is a decrease of the evacuation time with
  increasing group size.}


\section{Introduction}
\label{sec:1}
The influence of social groups in pedestrian dynamics, especially in
evacuation scenarios, is an area of recent interest, see e.g.
\cite{mueller2,mueller} and other contributions in these proceedings.
The situations that are considered are widespread and well-known in
everyday life. For example, many people visit concerts or soccer
matches not alone, but together with family and friends in so-called
social groups. In case of emergency, these groups will try to stay
together during an evacuation. The strength of this cohesion depends
on the composition of the social group. Several adult friends would
form a loose group that is mainly connected via eye contact, whereas a
mother would take her child's hand and form a strong or even fixed bond.
In addition, even the size of the social groups could have an effect
on the evacuation behaviour.

In order to consider these phenomena in a more detailed way, a
cooperation of researchers of the universities of Cologne and
Wuppertal and the Forschungszentrum J\"ulich has performed several
experiments aiming at the determination of the general influence of
inhomogeneities on pedestrian dynamics. They contained two series of
experiments with pupils of different ages in two schools in Wuppertal.
The first series focussed on the determination
of the fundamental diagram of inhomogeneous groups, i.e. pedestrians
of different size.
The second series of experiments considered evacuation scenarios. In
several runs the parameters of the crowd of evacuating pupils were
varied, i.e. the size of the social group and its structure 
and the interaction between the group members. 
Here we present first results for these evacuation experiments.


\section{Teaching units}
\label{sec:2}
The experiments were accompanied by teaching units for all involved
students providing an introduction into the topic of traffic
and pedestrian dynamics.

\begin{figure}[t]
  \sidecaption
\includegraphics[scale=.045]{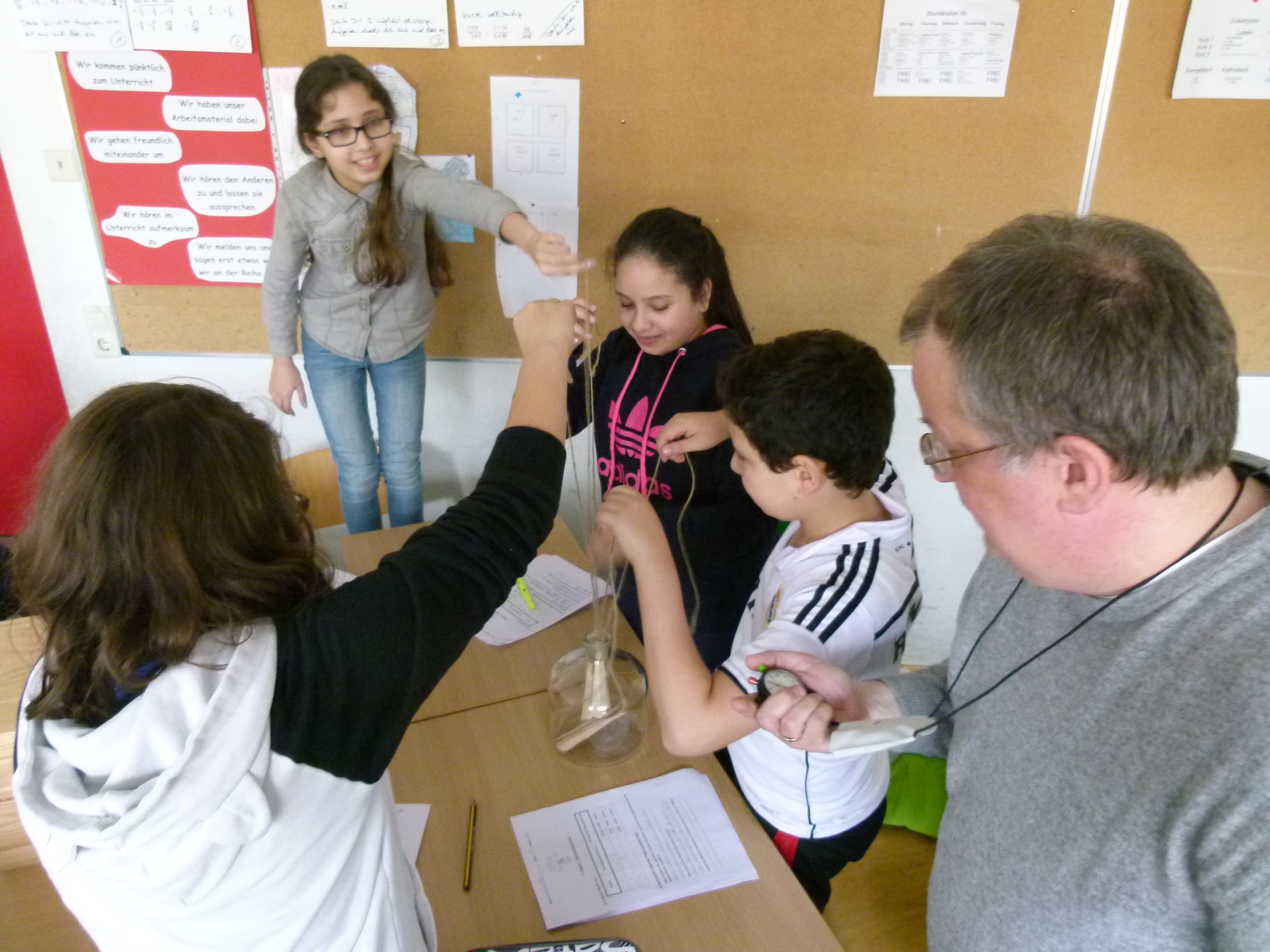}
\caption{Panic experiment according to Mintz. Every pupil is 
  assigned a cord with a wedge on its end lying in a bottle with a
  narrow neck. If every student pulls at the same time and as fast as
  possible, the wedges block at the bottleneck. On the other hand,
  behaving in a coordinate way leads to a smooth process that is
  significantly faster. (Photo by V.~Ziemer, U Wuppertal)}
\label{fig:1}       
\end{figure}

In classes of fifth and sixth grade, the focus of the classes was on
the important quantities of pedestrian dynamics, especially density,
time and bottleneck situations.  This introduction to crowd effects
and pedestrian behaviour was intended to raise awareness for their
relevance for their everyday lives and safety issues. Therefore we
arranged little experiments the students could perform themselves,
e.g. the panic experiment according to Mintz \cite{mintz} (see
Fig.~\ref{fig:1}). In small groups the pupils had to pull several
wooden wedges out of a bottle with a narrow neck as fast as possible
and observe the blocking of the wedges when every students pulls at
the same time. This experiment was supposed to indicate that
coordination can lead to better results compared to selfish behavior.

The older pupils of classes 10 and 11 participated in an introduction
to cellular automata and the physics of traffic. They received several
worksheets on the Game of Life and other cellular automata,
especially the Nagel-Schreckenberg model \cite{nasch}.  The aim of
these lessons was to obtain a first qualitative and quantitative
understanding of the collective effects in traffic systems.  This
should help to increase the identification with the experiments they
later participated in and raise awareness about the relevance of this
kind of research for everyday life.

\section{Experimental set-up and procedure}
\label{sec:3}
The experiments were performed in two schools in Wuppertal. Students of
four classes participated as part of project work. The experimental
room was built in the school's assembly hall.


\subsection{Experimental set-up}

The experimental area was a square room of $5\times 5$~m$^2$ bounded
by several small buckets. In the center of this area there was a
square starting area of $3\times 3$~m$^2$ denoted by the white marks.
The students stepped into the room through the entrance that is shown
below in Fig.~\ref{fig:2} and assembled in the starting area. During
the evacuation they had to leave the room using the exit on the left
side.  The exit door was built by two upstanding platforms and had a
variable width changing between 0.8~m and 1.2~m. The area behind the
door was connected to the waiting area before the entrance, so the
students could walk on a closed path.  For the collection of data all
experiments were recorded by a camera system.  This system was mounted
on the hall's ceiling and contained customary digital cameras and GoPros.

All students wore caps of different colour.  Each colour represented a
certain interval of body heights. The body height of each pupil was
measured before the experiments started. This information is needed
to determine the position accurately, but the different colours can 
also be used to draw conclusions about the composition of the group of
evacuating pupils later in the video. All caps showed also a black
point at the middle of the head. That allows to recognize and track
each person in the video.

\begin{figure}[t]
\sidecaption
\includegraphics[scale=.13]{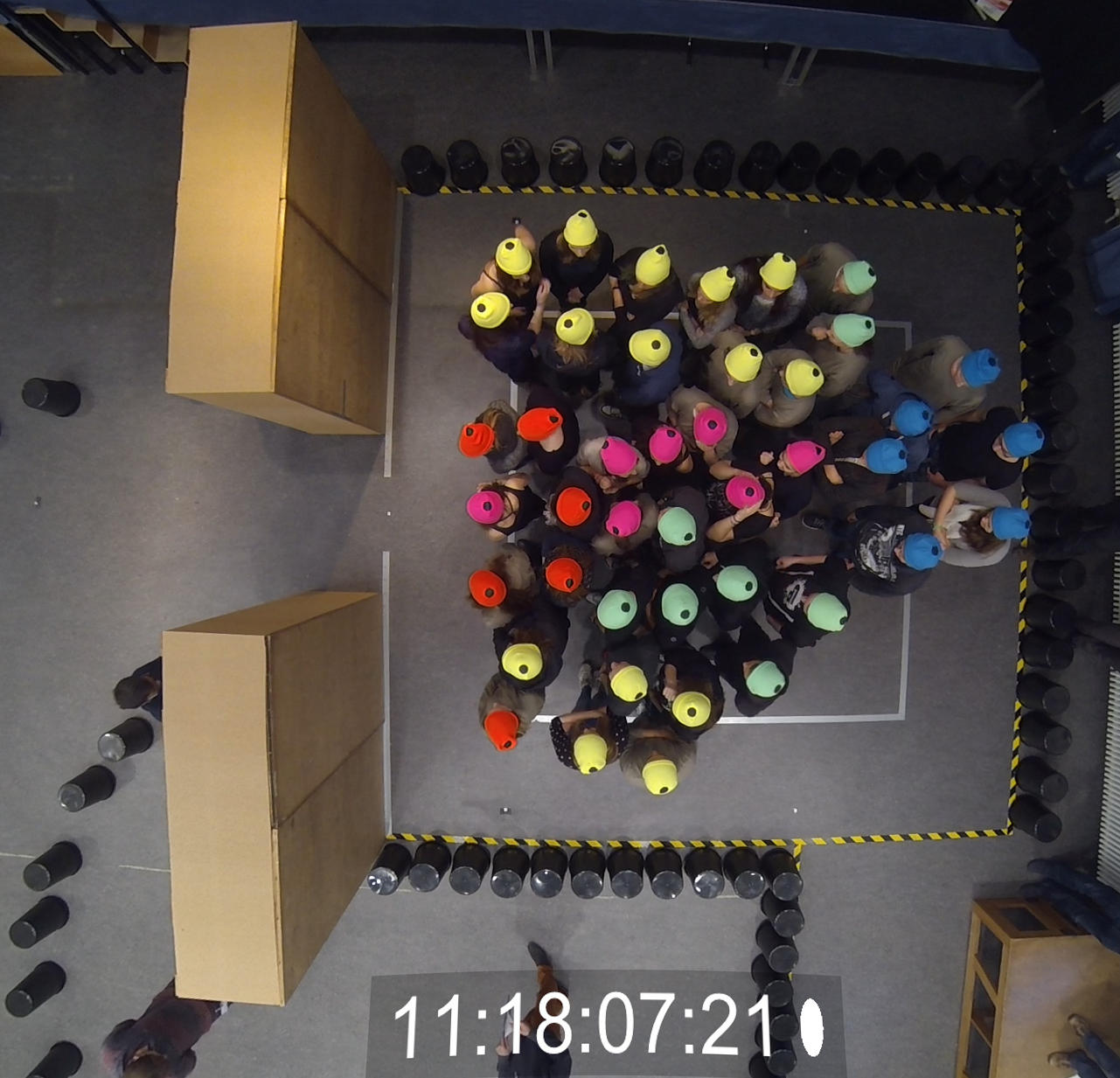}
\caption{The experimental area consists of a square room of $5\times 5$~m$^2$, 
  with a starting area in the middle. The pupils had to leave this
  room through the exit at the left. They wore colourful caps to
  distinguish the different body heights.}
\label{fig:2}       
\end{figure}


\subsection{Experimental procedure}

In general, the pupils had to perform several evacuation runs. For
each run, a group of 32 to 46 persons assembled in the starting area,
distributed nearly uniformly.  During the evacuation, the students
were allowed to use the whole experimental area.

After a starting signal, the participants had to leave the room using
the exit door. They should walk briskly and evacuate the room as fast
as possible. The pupils were told to imagine there would be a kind of
danger, like fire or smoke. However, they were not allowed to run,
scramble or push each other. After leaving the room they had to
assemble again in the waiting area in front of the entrance and to
wait for the next run.

The group of pupils that was placed into the experiment was compound
in different ways to consider different parameters.

The first parameter that was varied in the experiment was the
composition of the entire group. At all, there were two different age
classes allowing for three different group compositions. The crowd
could consist only of children aged 10 to 12 years, only of young
adults aged 15 to 17, or a mixture of both groups whereby children and
youths were equally represented.

The second parameter concerned the social group size. In several runs,
the students had either to evacuate on their own without regarding the
others around them, or to form pairs, or larger social groups. These
groups could contain four, six or eight persons. Within one pair or
social group the students had to try to stay together during the
evacuation run.

As a third parameter we considered the interaction within the social
group. The nature of the interaction can be specified by (i) its
strength and (ii) the hierarchy of group members. Regarding the
interaction strength, the group members could either be connected
loosely, by just trying to stay together via eye contact, or they
could have a fixed bond. A fixed bond was realized by holding each
other's hand or some other physical contact. Furthermore, hierarchy of
the group members could be different. In the first case, all partners
were treated equally. Each group member had to leave the room and to
stay together with their partners. In the other case, one student was
declared as the "leader", the other one as the "follower". The leader
had to leave the room without regarding its partner or the other
students, whereas the follower just had to follow the leader through
the room.

This leads to four different ways to form pairs during the evacuation
run. In the case of age-matched partners, the leader was chosen
randomly. For the runs with the mixed crowd, the pairs were composed
of one child and one teenager that took the part of the leader in the
runs they were needed. All runs with larger social groups were done
with loose bonds. In social groups of same age, there was no leader,
but in mixed social groups one of the youths was declared as the
leader.

To analyze the experiments the videos of the camera system that was
mounted on the hall's ceiling was available. For each run of the
experiment there is a video sequence. Using the \texttt{PeTrack}
software \cite{boltes}, it is possible to extract the trajectories for
each person and each run. The students were recognized via the black
point in the middle of their coloured caps. The position of this point
was tracked in each frame, generating the trajectory of each
participant.


\section{Analysis}
\label{sec:4}
First, we focus on the analysis of the data regarding the influence of group
size on the evacuation scenario. Therefore, we use the data of one
school and of the runs with larger groups. Most of these experiments
were performed only with the older pupils with loose bonds and no
leader-follower relationship, to which we restrict our analysis for
now.

In different runs, the students formed groups of four, six and eight
persons. In addition, one run with groups of six students and with an
explicitly cooperative behaviour within the group was performed. They
should concentrate a bit more on their group members and try to leave
the room together. For comparison, we also consider the run with pairs
and a loose bond that can be seen as a smaller group of two persons.


\subsection{Evacuation times for large groups}
\begin{figure}[t]
\sidecaption
\includegraphics[width=\textwidth]{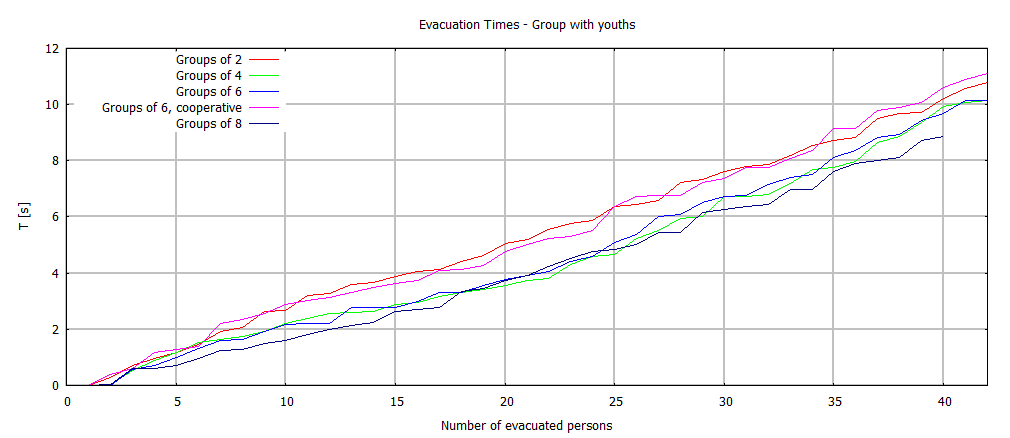}
\caption{Evacuation times for larger groups of one of the schools. 
  The splitting of the curves into two groups is obvious. The shorter
  evacuation times belong to the runs with larger groups of four, six
  and eight persons, the longer ones to those with pairs and
  cooperative behaviour.}
\label{fig:3}       
\end{figure}

First we consider the evacuation time. In Fig.~\ref{fig:3}, we plot
the number of evacuated persons against the time needed to leave the
room. The results can be compared between the different runs.

The evacuation time for each person is defined as the time difference
between the beginning of the evacuation and the moment when the person
passes the door, exactly when he/she leaves the aisle that is formed
by the two platforms. The beginning of the evacuation can be
determined only approximately because the starting signal is not
audible in the videos that are used for the extraction of the
trajectories. For extracting the evacuation times we set the beginning
on the moment of the first movement towards the door. However, for the
analysis the influences of this inaccurate definition, the
pre-movement time or other delays should be minimized. In doing so, we
take the evacuation time of the very first person that left the room
and subtract it from all the other times. Thereby, all plots start at
zero for the first person and it is easier to compare different runs.

For the analysis of the runs with larger groups the evacuation times
are shown in Fig.~\ref{fig:3}. All graphs show a nearly linear
behaviour that could be expected. At the beginning of the evacuation
all evacuation times are roughly the same. Between three and six
evacuated persons the curves start to split into two groups. After
increasing slightly, the difference between the two progresses remains
nearly constant until the end of the evacuation. The main insight is
that there are several runs that are clearly faster than other ones.

The upper two curves represent the evacuation in pairs and in groups
of six with very cooperative behaviour. The lower graphs show the runs
with larger groups of four, six and eight persons. Within the two
groups of curves the differences are not large enough to separate the
runs from each other. However, in the lower group, the run with eight
participants per group seems to be a bit faster at the end of the
evacuation. The run with six participants per group and cooperative
behaviour is clearly slower than the run with same group size but
without this instruction. These results indicate that forming groups
is advantageous for the evacuation, whereas behaving cooperatively
inhibits this effect.

While looking for reasons for the differences in evacuation times, one
first approach could be to determine the density distribution.
Therefore, we determined the Voronoi cells \cite{steffen,voronoi}
within the experimental room for each person at different times. As a
measure of density we coloured all cells dependent on their size:
smaller cells are coloured in shades of red, larger ones in blue.

\begin{figure}[t]
\begin{center}
\sidecaption
\subfloat[Pairs at 4 s]{\includegraphics[scale=.30]{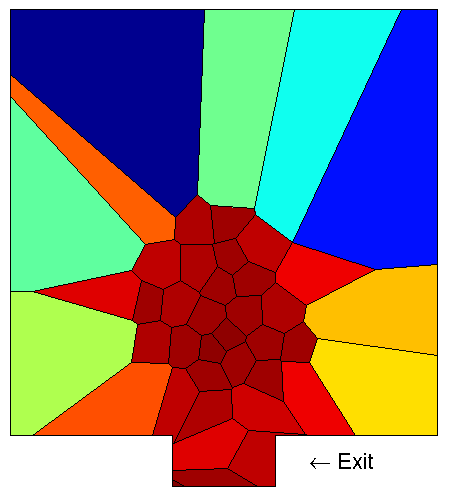}}
\qquad
\subfloat[Groups of four at 4 s]{\includegraphics[scale=.30]{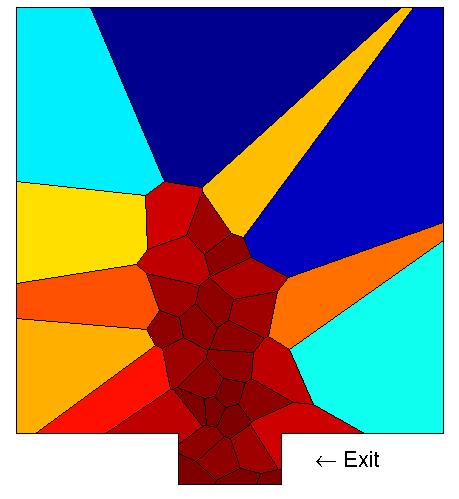}}
\caption{Voronoi cells for each person in the room at time 4~s 
  for the runs with pairs and social groups of four persons. The
  colour of each cell depends on its size as a measure of density.}
\label{fig:4}
\end{center}       
\end{figure}

In Fig.~\ref{fig:4} the density distributions for the run with pairs
and with groups of four persons are shown. It is clearly seen that the
distribution for the run in (b) is a bit narrower than the other one
at the same time step. That means when forming groups, the children
order rather behind each other than next to each other in front of the
door. This behaviour seems to be advantageous for evacuating the room
as it leads to a shorter evacuation time.

\subsection{First attempts to interpretation}
\label{sec:5}
The results obtained so far suggest certain interpretations which,
however, need to be substantiated by further experiments with better
statistics. It is obvious from the plot of the evacuation times that
increasing the group size leads to a decrease in evacuation times. The
density distributions show the pupils ordered in different ways for
forming groups than for pairs. A possible explanation is that the
persons subordinate within the group and just follow the other group
members. Because of that, there may be less conflicts between persons
that meet at the door in competing for space. A person is just in
competition with persons of other groups, not with own group members.
Increasing the group size reduces the number of possible competitors.
This reduction of conflicts may have a positive influence on the
evacuation time.

When the children have to show cooperative behaviour, the evacuation
is slower than without this instruction. It is a possible explanation
that here the effort to stay together is larger and reduces the effort
to leave the room.


\section{Summary and outlook}
\label{sec:6}
We performed experiments under laboratory conditions to determine the
influence of social groups on evacuations. A comparison of evacuation
times between runs with different group sizes shows that increasing
the group sizes lowers the evacuation time. The participants order in
a different way for larger groups.  

These first preliminary results have to be analyzed in more detail.
The statistics need to be improved by further experiments.  However,
with the help of the density distributions, photographs of the finish
and the data of the second school we hope to get more information from
the present experiments, e.g. about the microscopic mechanisms
especially close to the exit.  In addition, there are some few
parameters that should also be analyzed, e.g. the effect of body size
and age.

\begin{acknowledgement}
We dedicate this contribution in grateful memory to our friend and 
colleague Matthias Craesmeyer.\\
We thank the team from the Forschungszentrum J\"ulich and Wuppertal
University and the students and teachers of Gymnasium Bayreuther
Stra\ss{}e and Wilhelm-D\"orpfeld-Gymnasium in Wuppertal for the help
with the experiments. Financial support by the DFG under grant 
SCHA~636/9-1 is gratefully acknowledged.
\end{acknowledgement}

%

\bibliographystyle{spphys.bst}
\bibliography{tgfbibs}

\end{document}